\documentclass[fleqn,usenatbib]{mnras}

\usepackage[T1]{fontenc}

\DeclareRobustCommand{\VAN}[3]{#2}
\let\VANthebibliography\thebibliography
\def\thebibliography{\DeclareRobustCommand{\VAN}[3]{##3}\VANthebibliography}

\usepackage{graphicx}	
\usepackage{amsmath}	
\usepackage{amssymb}	

\usepackage{newtxtext,newtxmath}

\title[The Connection between UV/Optical Variations and Radio Emission in Radio-Quiet Quasars]{Exploring the Connection between UV/Optical Variations and Radio Emission in Radio-Quiet Quasars: Clues on Radio Emission Origin}

\author[M. Liao et al.]{
Mai Liao,$^{1,2}$\thanks{E-mail: liaomai@ustc.edu.cn}
Junwan Wang,$^{1,2}$\thanks{E-mail: jxw@ustc.edu.cn}
Wenyong Kang$^{1,2}$
Minhua Zhou$^{3}$
\\
$^{1}$CAS Key Laboratory for Research in Galaxies and Cosmology, Department of Astronomy, University of Science and Technology of China, Hefei, Anhui 230026, \\
China\\
$^{2}$School of Astronomy and Space Science, University of Science and Technology of China, Hefei 230026, China\\
$^{3}$School of Physics and Electronic Information, Shangrao Normal University, 401 Zhimin Road, Shangrao 334001, China
}

\date{Accepted 2022 January 26; Received 2021 December 30; in original form 2021 October 15}

\pubyear{2021}

\begin{document}
\label{firstpage}
\pagerange{\pageref{firstpage}--\pageref{lastpage}}
\maketitle

\begin{abstract}
The radio emission in radio-quiet quasars (RQQs) has been a long mystery and its physical origin remains unclear.
In a previous work we find UV/optical more variable quasars have stronger X-ray emission, indicating a link between disc turbulence and X-ray corona heating. In this work, for the first time, we investigate the relation between UV/optical variability and the radio emission in RQQs selected from SDSS stripe 82 and FIRST radio survey. 
We median stack the FIRST images and detect clear signals from RQQs in the co-added images of individually radio non-detected sources.
Controlling the effects of other parameters, including redshift, black hole mass, bolometric luminosity and Eddington ratio,
we find more variable RQQs, which are known to be X-ray relatively brighter, show tentatively weaker radio emission, contrary to the linear X-ray/radio correlation if the radio emission is from or driven by the corona. The discovery also suggests that if the radio emission in RQQs is driven by AGN activity (such as weak jet), the underlying driving process is independent to the disc turbulence which drives UV/optical variability and probably also corona heating. Alternatively, the radio emission could be due to star formation in the host galaxies.
\end{abstract}




\section{Introduction}

Quasars, powered by the central accreting supermassive black holes, emit light throughout the entire electromagnetic spectrum from radio to X-rays and $\gamma$-rays. 
Traditionally, based on the radio emission loudness which is defined as the ratio between radio and optical flux (e.g., $R = f_{\rm 5GHz}/f_{\rm 4400\AA}$, \citealt{1989AJ.....98.1195K}), quasars could be divided into the two populations of radio-loud (RL) and radio-quiet (RQ) \citep[e.g.,][]{2017A&ARv..25....2P, 2019NatAs...3..387P}. But see \cite{Padovani2017NA} for an alternative naming of RL/RQ as ‘jetted’ and ‘non-jetted'.
Powerful relativistic and well collimated jets \citep[e.g.,][]{2019ARA&A..57..467B,2020NewAR..8801539H} are commonly presented in radio-loud quasars (RLQs), which account only approximately 10-20$\%$ \citep[e.g.,][]{1989AJ.....98.1195K,2002AJ....124.2364I} of the whole population.
The radio emission of the rest majority (i.e., radio-quiet quasars, RQQs) is typically 1000 times fainter \citep[e.g.,][]{1993MNRAS.263..425M,2019NatAs...3..387P},
the origin of which is still under debate. 
Possible physical origins of the radio emission in RQQs include low-power compact jets, accretion-disc corona, outflows, and star formation in the host galaxies \citep[e.g.,][]{Padovani2016,2019NatAs...3..387P}. 

Resolving the expected extended, diffuse radio emission associated with star formation in the host galaxies of distant RQQs is much more challenging than that in nearby lower luminosity radio-quiet AGN \citep[e.g., VLA and eMERLIN observations of NGC 1614,][]{2010A&A...513A..11O}. 
The commonly detected compact radio structures on arcsecond scales in RQQs \citep[e.g.,][]{2005ApJ...621..123U,2006A&A...455..161L,2020MNRAS.492.4216S} are often insufficient to separate radio emission originating from star-formation and AGN-related activity.
Comparing the radio power in RQQs with the star formation activities in their host galaxies also yielded mixed results, with some favoring the star formation origin \citep[e.g.][]{2015MNRAS.453.1079B,2016ApJ...831..168K,2019A&A...622A..11G} and others favoring black hole activity \citep[e.g.][]{2016MNRAS.455.4191Z, White2017}.

AGN driven radio emission in RQQs could be immediately confirmed if the linear jet-like structures, which may be a scaled-down version of the more powerful ones in RLQs \citep[e.g.,][]{1995A&A...293..665F,2019NatAs...3..387P,2021MNRAS.506.5888M}, 
could be directly resolved
\citep[e.g.,][]{1994AJ....108.1163K,1998MNRAS.297..366K,2001ApJ...562L...5B,2003ApJ...591L.103B,2005ApJ...621..123U,2006A&A...455..161L,2021MNRAS.504.3823W}. The observed flat radio spectra and the 
high-brightness temperature ($> 10^7$ K) of the compact radio cores also favor the non-thermal processes from relativistic electrons usually related with jets \citep[e.g.,][]{1979ApJ...232...34B,1998MNRAS.299..165B,2005ApJ...621..123U,2013MNRAS.432.1138P}. However,
although lower power jets have been detected in a handful of RQQs, 
most RQQs studies with high-resolution radio observations show unresolved radio structures, even at milli-arcsecond scales by VLBI imaging \citep[e.g.,][]{2005ApJ...621..123U}.
The uncollimated jets, i.e., outflows, might also contribute to the radio emission in RQQs. They may produce radio emission with non-isotropic, conical and spatially extended structures \citep{2019NatAs...3..387P}, and has been proposed to explain the observed correlation between the velocity width of [O III] and the radio emission in RQQs \citep{2013MNRAS.433..622M,2014MNRAS.442..784Z,2018MNRAS.477..830H}. 

Accretion-disk corona is also an option for radio emission origin of RQQs \citep{2008MNRAS.390..847L,2016MNRAS.459.2082R}, as the evidence showed in \cite{2008MNRAS.390..847L} that 71 radio-quiet PG quasars follow the  well-established relation of $L_{\rm R_{5GHz}} / L_{\rm X} \sim 10^{-5}$ in coronally active cool stars \citep{Guedel1993}. 
Meanwhile, the detections of millimeter excess emission (100-300 GHz) in a few local Seyfert galaxies suggest the existence of a compact, optically thick core, which could be associated with the corona \citep{2018MNRAS.478..399B,Inoue2018}. 
Recently it was found X-ray luminosity of radio-quiet PG quasars correlate more tightly with the 45-GHz luminosity than the 5-GHz \citep{Baldi2021}, suggesting the 45-GHz emission originates probably from the accretion disc corona.
High-frequency observations for more RQQs however are challenging due to the weakness of the signal. Furthermore, studies on the correlation between X-ray and radio emission (at any frequency) in larger samples of RQQs are scarce. 

In this work we explore the physical origin of radio emission in RQQs from a new perspective, i.e., to study the relation between radio emission and UV/optical variability in a large sample of RQQs.
Aperiodic variability from radio to $\gamma$-ray, over a wide range of timescales from minutes to years, is one of the prominent characteristics of quasars \citep{2017A&ARv..25....2P}. In UV/optical, 
the variability is generally attributed to random fluctuations in the accretion disc, probably driven by magnetic turbulence \citep[e.g.][]{2009ApJ...698..895K,2016ApJ...826....7C,2018ApJ...855..117C,2018ApJ...868...58K}. 
Remarkably, \cite{2018ApJ...868...58K} found a intrinsic link between X-ray loudness (defined as $L_{\rm X}/L_{\rm bol}$) and UV/optical variability amplitude in normal quasars,  i.e., more variable quasars have stronger X-ray emission, suggesting a physical link between disk turbulence and corona heating. 
Thus exploring the connection between UV/optical variability and radio emission enables us to at the same time study the connection of X-ray and radio emission in a large sample of RQQs.
If radio emission in RQQs is dominated by the corona activity, we thus expect more variable quasars should also have stronger radio emission. 
Furthermore, the UV/optical variability is a feature intrinsic to the accretion process. Examining 
the correlation between radio emission and UV/optical variability is useful to determine whether the radio emission is associated with such nuclear activities. Such correlation, if exists, could yield new clues to the understanding of radio emission origin. 

In this paper, we explore such connection using SDSS stripe 82 quasars within FIRST survey footprint.
We stack FIRST images of quasars to measure the average radio emission.
The outline of our work is as follows: \S2 presents the sample and the stacking analyses of radio images; \S3 presents our results; \S4 provides a discussion. Throughout the paper. the cosmological parameters $H_0 = 70\, \mathrm{km\,s^{-1}\, Mpc^{-1}}$, $\Omega_\mathrm{m} = 0.3$, and $\Omega_{\lambda} = 0.7$ are adopted.

\section{Sample and Stacking analysis}
\subsection{The quasar sample}
In order to build quasar sample with UV/optical variability measurements, we start from the SDSS Stripe 82 survey which covers about 290 $deg^2$ equatorial field of the sky and has been scanned over 60 times in the $ugriz$ bands by Sloan Digital Sky Survey \citep{2007AJ....134.2236S}.  \cite{2012ApJ...753..106M} provided the re-calibrated $\sim$10 years long light curves in $ugriz$ for 9258 spectroscopically confirmed quasars in SDSS Data Release 7 (SDSS DR7)\footnote{ http://faculty.washington.edu/ivezic/cmacleod/qso$\underline{~~}$dr7/Southern.html}. 9120 out of them 
are located within the footprint of the Faint Images of the Radio Sky at Twenty-Centimeters (FIRST) survey, which is conducted with NRAO Very Large Array (VLA) in its B-configuration centered at 1.4 GHz and covers about 10,575 square degrees\footnote{http://sundog.stsci.edu} \citep{1995ApJ...450..559B,2015ApJ...801...26H}. 
The FIRST images have a resolution of $\sim$ 5$''$, and a typical rms sensitivity of 0.15 mJy.
We cross-match with the FIRST catalog \citep{2015ApJ...801...26H} with a matching radius of 5$\arcsec$, and find 517 (5.7\%) sources with FIRST counterparts.
Among the FIRST detected sources, $\sim$ 80\% could be classified as radio-loud based on their radio-loudnesses.
This indicates while the dominant population of the sample is radio-quiet, some ($\sim$ 6\%, assuming $\sim$ 10\% of the 9120 quasars are radio-loud) radio non-detected sources could still be radio-loud.

We further match the quasars with \cite{2011ApJS..194...45S} to obtain their $z$, $M_{\rm BH}$, $L_{\rm bol}$ and $R_{\rm edd}$, where we adopt the fiducial virial $M_{\rm BH}$ from their catalog. 
108 objects are further rejected due to the lack of $M_{\rm BH}$.

We adopt the same method as in \cite{2018ApJ...868...58K} to measure the intrinsic variability amplitude in each photometric band for each quasar. The variation amplitude is defined as excess variance $\sigma_{\rm rms}$ \citep[e.g.,][]{2007AJ....134.2236S,2012ApJ...758..104Z} as follows:

\begin{equation}
  \sigma_{\rm rms}^2=\frac{1}{N-1}\sum(X_i-\bar{X})^2 - \frac{1}{N}\sum\sigma_i^2
\end{equation}

where $N$ is the number of photometric measurements for a single light curve, $X_i$ the magnitude for $i_{\rm th}$ observation, $\bar{X}$ the average magnitude, and $\sigma_i$ the $i_{\rm th}$ photometric uncertainty of each observation. We exclude $u$ and $z$ band data due to their significantly larger photometric uncertainties when compared with other three bands of $gri$. With only keeping the ones which have at least 20 epochs in each of the light curve to ensure accurate measurement of $\sigma_{\rm rms}$, our final sample consists of 8819, 8829, 8826 RQQs with UV/optical variability amplitude in $g$, $r$, $i$ band, respectively. 
Below we focus on $g$ band data as quasars are generally more variable in $g$ band than in $r$ or $i$ band. Unless otherwise stated, we present below the results using $g$ band $\sigma_{\rm rms}$, which are further verified by the well consistent results obtained using $r$ or $i$ band data.

\subsection{High-/low-variability subsamples}

In order to study the relation between radio emission and UV/optical variations in RQQs, we firstly divide our quasars equally into two subsamples according to $\sigma_{\rm rms}$, that is, a high-variability and low-variability subsample (hereafter HVQ0 and LVQ0 respectively). 

\begin{figure}
	\includegraphics[width=1.05\columnwidth]{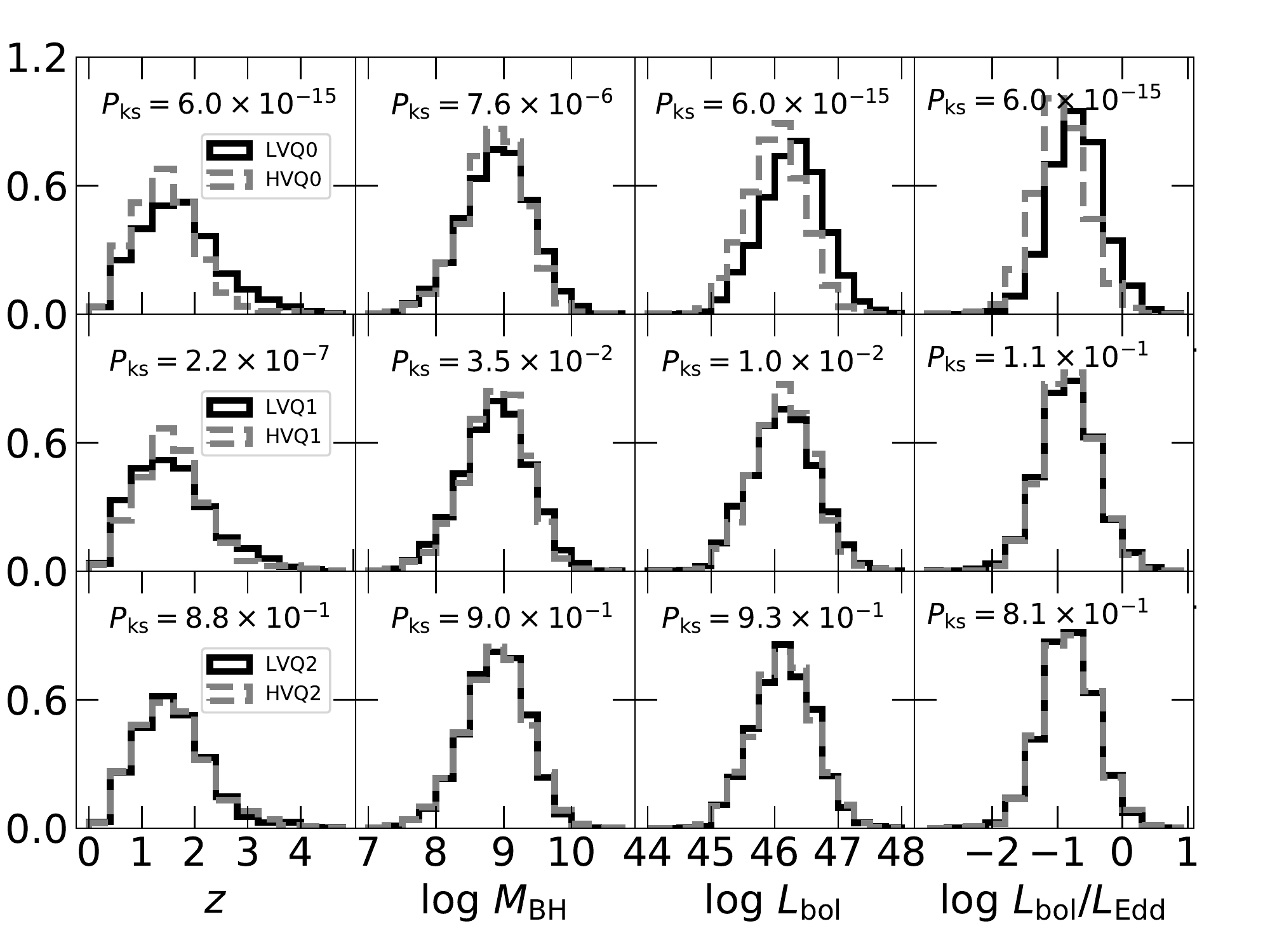}
	\caption{The distributions of $z$, $M_{\rm BH}$, $L_{\rm bol}$ and $R_{\rm edd}$ for HVQ0/LVQ0, HVQ1/LVQ1, HVQ2/HVQ2 from upper to lower panels, respectively. The K-S test results between the distributions are given in the plot.  \label{distribution_no_match}}
\end{figure}

However it is known that the UV/optical variability amplitude of quasars depends on physical parameters including luminosity, SMBH mass, Eddington ratio, and redshift \citep[e.g. see][ and references therein]{2018ApJ...868...58K,2021ApJ...911..148K}. Indeed, according to Kolmogorov-Smirnov (K-S) test, the HVQ0 and LVQ0 subsamples we obtained above do have significantly different distributions in redshift, bolometric luminosity, SMBH mass and Eddington ratio, and LVQ0s have on average higher redshift, luminosity, SMBH mass and Eddington ratio (see Fig. \ref{distribution_no_match}).
The radio flux in RQQs could also be sensitive to these parameters. A fair comparison to reveal the intrinsic connection between UV/optical variability and radio emission in RQQs should be made with these parameters matched.

\begin{figure}
	\includegraphics[width=1.0 \columnwidth]{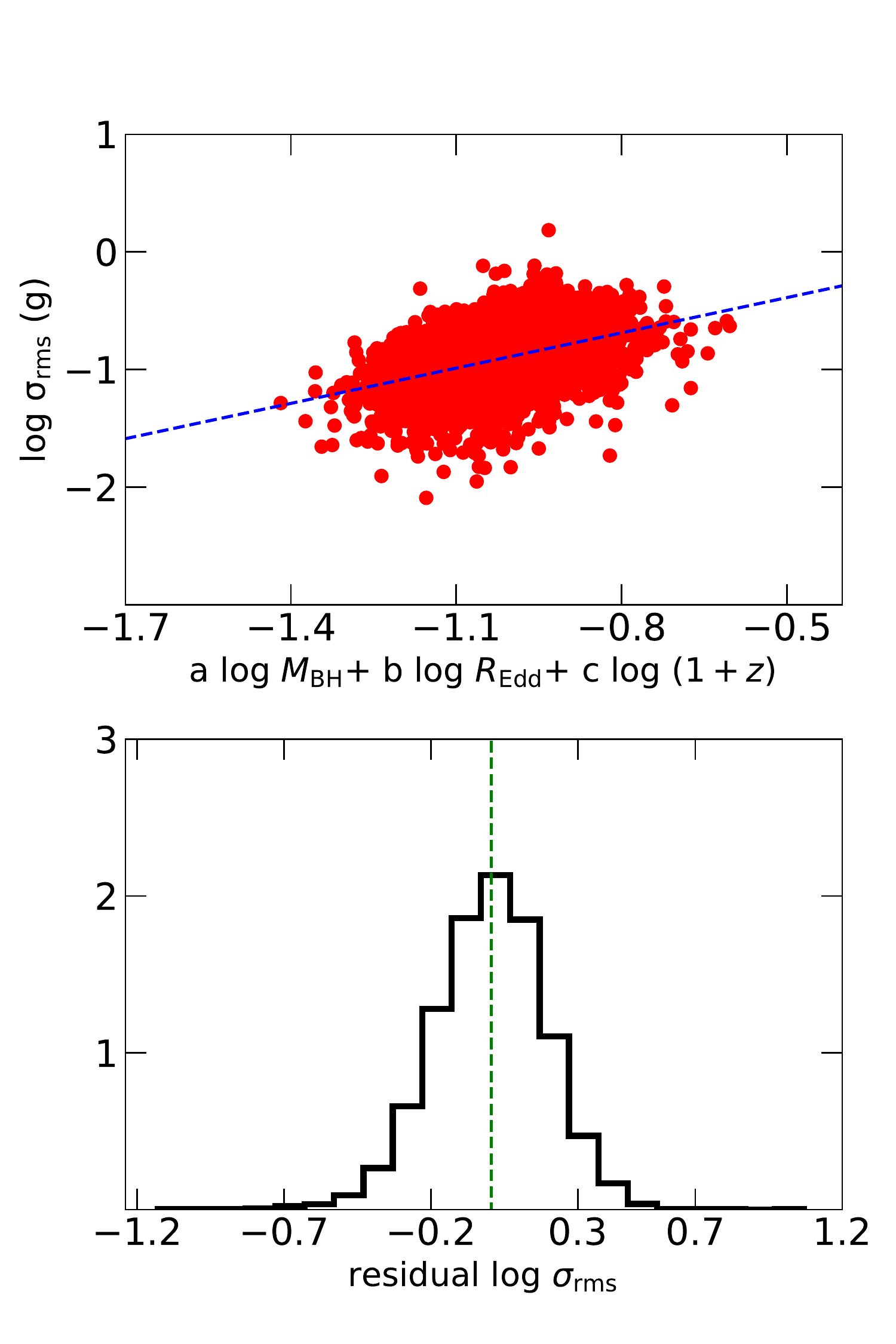}
	\caption{Upper: the observed SDSS $g$ band $\sigma_{\rm rms}$ versus the expected value based on the best-fit correlation (blue dashed line) in equation \ref{Eq-regression}. Lower: the distribution of the residual $\sigma_{\rm rms}$ (observed minus expected). The vertical green dashed line marks the segmentation between HVQ1 and LVQ1.\label{rms_fit}}
\end{figure}

As the stacked signals would be too weak if we split the samples into too many small bins in the multidimensional space of these parameters, we adopt two different approaches to divide the parent sample again into just two subsamples but with the parameters aforementioned matched between them. 
We first follow \cite{2018ApJ...868...58K} to perform multiple linear regression as follows: 
\begin{equation}\label{Eq-regression}
   \sigma_{\rm rms} \sim M_{\rm BH}^{a}R_{\rm edd}^{b}(1+z)^{c}
\end{equation}
We derive the residual $\sigma_{\rm rms}$ for each source from above equation, i.e., the difference between the observed $\sigma_{\rm rms}$ and the expected one based on the best-fit regression (see Fig. \ref{rms_fit}). 
And we divide the parent sample again into low-variability quasars (hereafter LVQ1) and high-variability quasars (HVQ1) based on the residual $\sigma_{\rm rms}$. The HVQ1 and LVQ1 subsamples do show consistent distributions of bolometric luminosity, SMBH mass and Eddington ratio (see Fig. \ref{distribution_no_match}), however the match in their redshift distributions is poor, likely because the dependence of $\sigma_{\rm rms}$ on redshift is more intricate than Equation \ref{Eq-regression} could describe\footnote{Various lines which would shift into or out of SDSS $g$ band at certain redshifts could alter the values of $\sigma_{\rm rms}$ as the lines are less variable than the underlying continuum emission, leading to sharp dependence of $\sigma_{\rm rms}$ on redshift.}.

We then adopt a different approach to build HVQ and LVQ samples with better matched parameters.
We randomly choose as many as possible and equal number of HVQ1s and LVQ1s (defined according to residual $\sigma_{\rm rms}$ in Fig. \ref{rms_fit}) in each three-dimensional bin of $z$, $M_{\rm BH}$, and $L_{\rm bol}$\footnote{Here, using $R_{\rm edd}$ to replace $L_{\rm bol}$ would obtain same results.} which contain both HVQ1s and LVQ1s. We endeavor to adjust the bins sizes of $z$, $M_{\rm BH}$, and $L_{\rm bol}$ to maximize the number of sources selected. Finally, we could select $\sim$ 88$\%$ sources out of the parent sample into HVQ2 and LVQ2 subsamples. This is because with the grid selection approach when unequal numbers of LVQ1s and HVQ1s fall in a grid, we have to drop sources to select equal number of LVQ1s and HVQ1s \citep[e.g. see][]{2021RAA....21....4Z}. 
As expected, the HVQ2 and LVQ2 subsamples derived with this approach show statistically consistent distributions in $z$, $M_{\rm BH}$, and $L_{\rm bol}$ and $R_{\rm edd}$ (see Fig. \ref{distribution_no_match}).

\subsection{FIRST image stacking}
Below we describe the procedures we adopted to derive the stacked FIRST image of a sample of quasars.
We extract from FIRST survey the
radio cutout image for each quasar centred on the SDSS quasar position (with size of $30'' \times 30''$, pixel size 1.8$''$). As most of our quasars are individually non-detected by FIRST,  
we employ the method of image stacking, a technique for combining data from many individual sources in order to study their statistical properties which are below the detection limit for a particular survey \citep[e.g.,][]{2007ApJ...654...99W, 2009MNRAS.394..105G}, to obtain the average radio emission. 
In this work, we adopt the median 
stacking approach, which is insensitive to outliers (i.e., the tails of the underlying flux distribution) and more robust for non-Gaussian distributions when compared with mean stacking \citep{2007ApJ...654...99W}. We align and median stack the individual FIRST images on a pixel-by-pixel basis \citep{2021A&A...649L...9R}, using the python code of numpy.ma.median \citep{2011CSE....13b..22V}. 
The average radio flux density of a sample could then be derived from the stacked image.
As a point-like source is centered at the nominal quasar(s’) position in each co-added image (see \S3) and the changing PSFs from various individual FIRST images in the stacking are not CLEANed\footnote{Although the most commonly used method to restore the flux density of a radio source from dirty image is CLEAN, maximum entropy and compressed sensing \citep{2017isra.book.....T}, the flux density of a radio source can also be derived from the peak flux density in dirty beam when the source is very compact (i.e. point source). 
Since our RQQs are individually non-detected, their CLEANed images are not available.
To avoid the problems caused by the fact that the PSF in the median stacked image is some kind of combination of various individual PSFs and it is not a CLEAN Gaussian, it is simpler to just use the value of central pixel as the peak flux density (Richard L. White, private communication).}, following \cite{2007ApJ...654...99W}, we use peak flux density only from the central pixel in each co-added image.

Another advantage of median stacking, when compared with mean stacking, is that the average radio flux derived from median stacking 
is less sensitive to the small population of radio-loud sources, and could better represent the dominant radio-quiet population.
Note the FIRST images are not sufficiently deep to detect/exclude all radio-loud quasars from our sample. 
To further minimize the effect of the small population of FIRST non-detected but radio-loud sources, we derive the 45th percentile\footnote{For each pixel in FIRST images to be stacked, we rank the flux densities in the pixel from all quasars in a sample, drop the 10\% with largest flux densities, and derive the median for the rest 90\%. } instead of the simple median (50th percentile) while stacking FIRST images, to reduce the effect of the $\sim$ 10\% radio-loud population.
This is practically valid because some radio-loud sources, though non-detected, may have marginal/tentative signals in FIRST images, and thus the radio-loud fraction is expected to be even lower (or negligible) in sources with apparent FIRST pixel values in the bottom 45\%. 
Note this approach (45th percentile) could only yield perfect measurement of average radio flux for RQQs (free from contamination by radio-loud ones) if the radio-loud fraction is 10\% in the whole sample, and none of the sources with FIRST pixel values in the bottom 45\% is radio-loud.
Hereafter we refer the 45th percentile of the whole sample (including those FIRST detected) as the median of RQQs, if not otherwise specified.

We finally stress that directly adopting the median of the whole sample
(instead of 45th percentile)
would yield slightly higher average radio flux densities, however, it won't alter the main results of this work. Using mean stacking wound not change the key results presented in this paper either.

\section{Results}
\begin{figure}
 	\includegraphics[width=1.0\columnwidth]{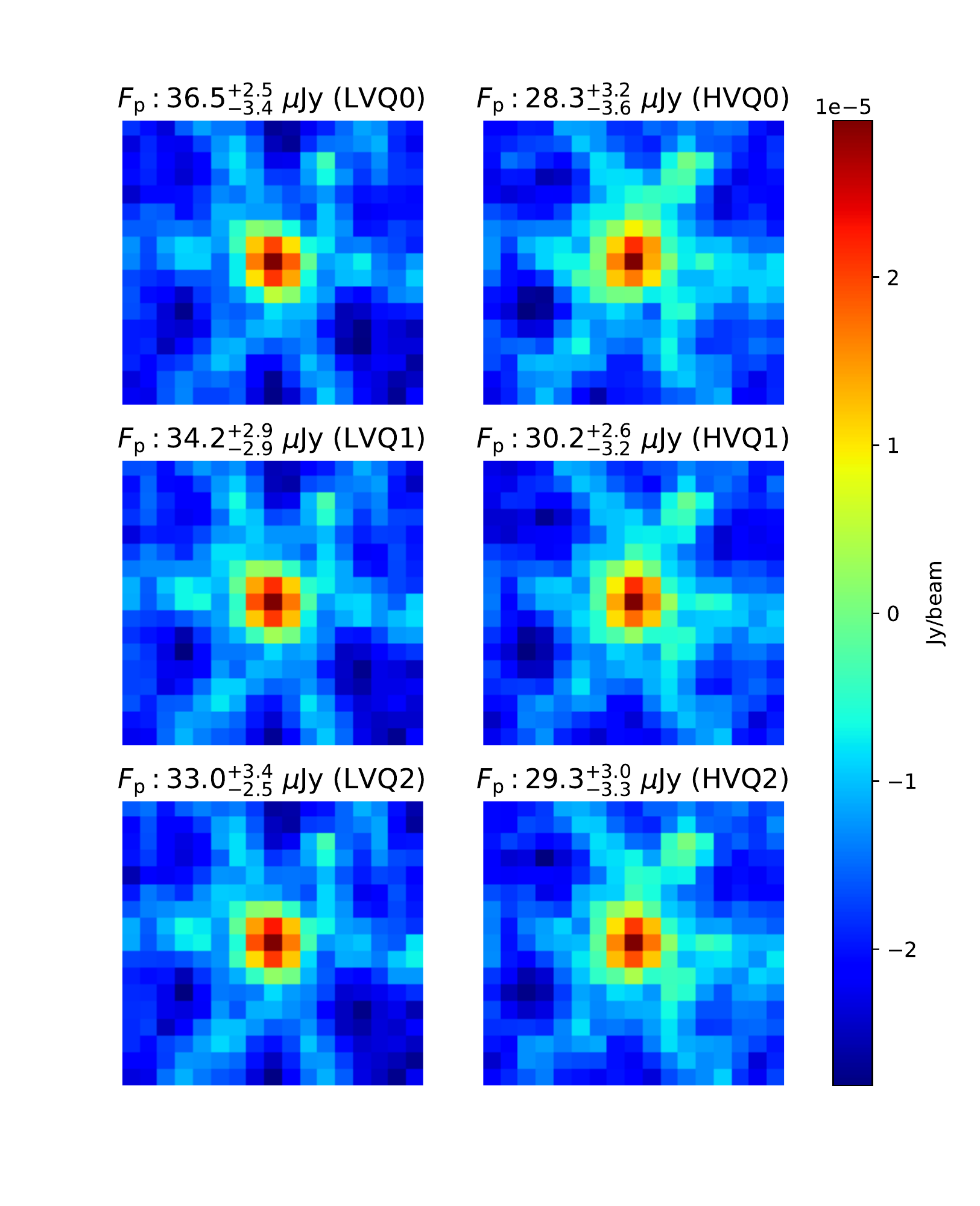}
	\caption{The co-added 1.4 GHz FIRST images of low-variability and high-variability RQQs (from upper to lower panels, respectively, for LVQ0s and HVQ0s, LVQ1 and HVQ1, and LVQ2 and HVQ2). 
	The peak flux densities indicated by the central pixel in each stacked image are shown at the top of each image. All the co-added images have a scale of $30'' \times 30''$ with pixel size 1.8$''$. The sidelobe effects are apparent due to the uncleaned PSFs.
	\label{stacked_image}}
\end{figure}

Fig \ref{stacked_image} shows the co-added FIRST images for H/LVQ (0/1/2). Clear compact radio signals are visible in the co-added images, showing the stacking approach is able to detect the average radio emission from our large sample of RQQs.
The asymmetric error bars for each peak flux density were derived using the bootstrapping technique described in \cite{2011ApJ...730...61K}.
It should be noted that our measurements of flux density values were not corrected for FIRST snapshot bias which exists for FIRST faint sources and our values above should be corrected by a multiplier of 
1.4 \citep[see details in][]{2007ApJ...654...99W}. As we aim to study the difference between subsamples, the snapshot bias would not affect our results.

\begin{figure}
	\includegraphics[width=1.0\columnwidth]{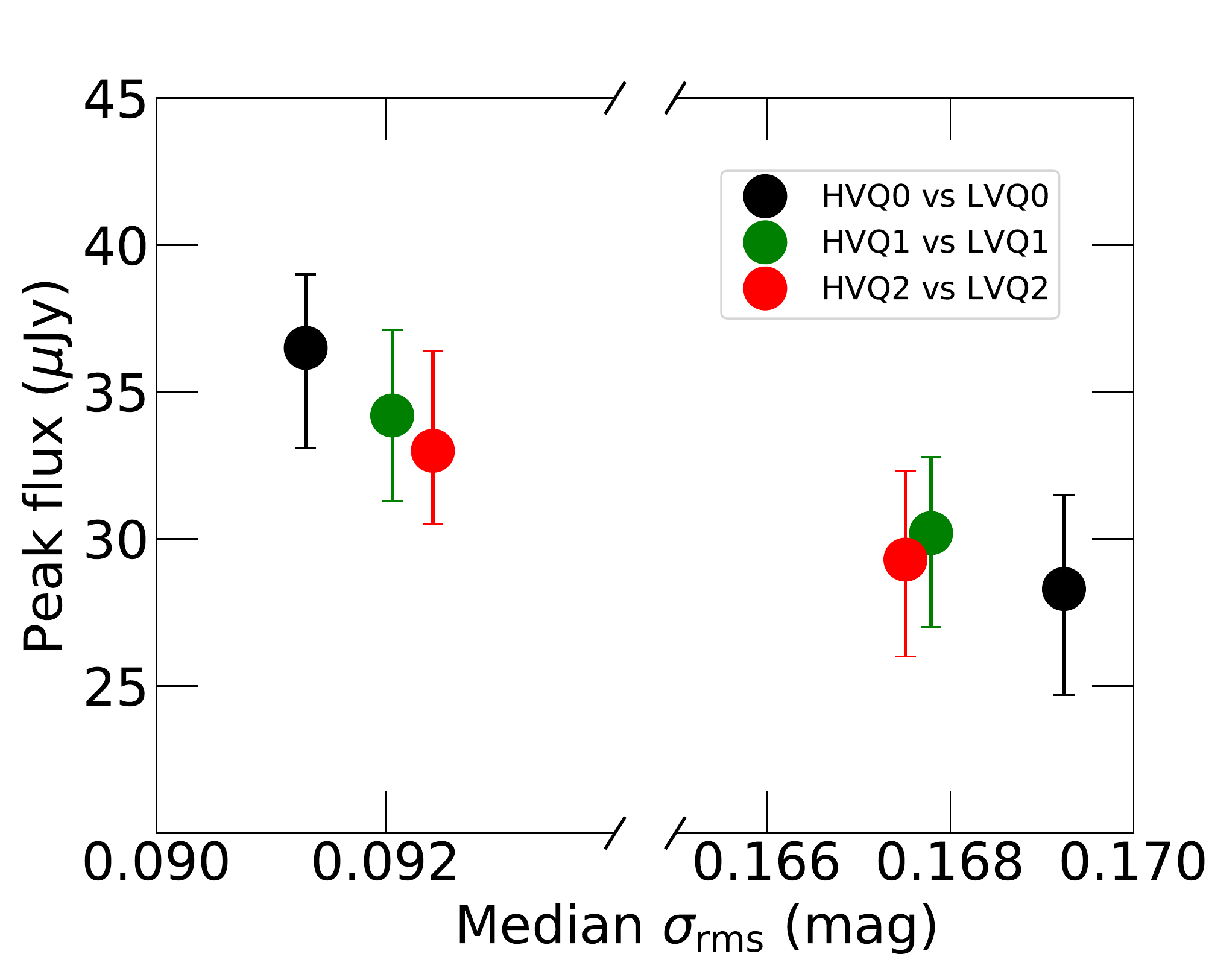}
	\caption{The median radio flux densities for HVQ0 and LVQ0 (black dots), HVQ1 and LVQ1 (green dots), and HVQ2 and LVQ2 (red dots). \label{bins5}}
\end{figure} 

The median-stacked flux densities of H/LVQ (0/1/2) are plotted in Fig. \ref{bins5}. The radio emission of the HVQ0 subsample is tentatively weaker (with a S/N of 1.7) compared with LVQ0 (28.3$_{-3.6}^{+3.2}$ $\mu$Jy vs 36.5$_{-3.4}^{+2.5}$ $\mu$Jy). Similar trends are seen in the comparisons between HVQ1 and LVQ1  (30.2$_{-3.2}^{+2.6}$ $\mu$Jy vs 34.2$_{-2.9}^{+2.9}$ $\mu$Jy), and between HVQ2 and LVQ2 (29.3$_{-3.3}^{+3.0}$ $\mu$Jy vs 33.0$_{-2.5}^{+3.4}$ $\mu$Jy).

\begin{figure}
	\includegraphics[width=3.4in]{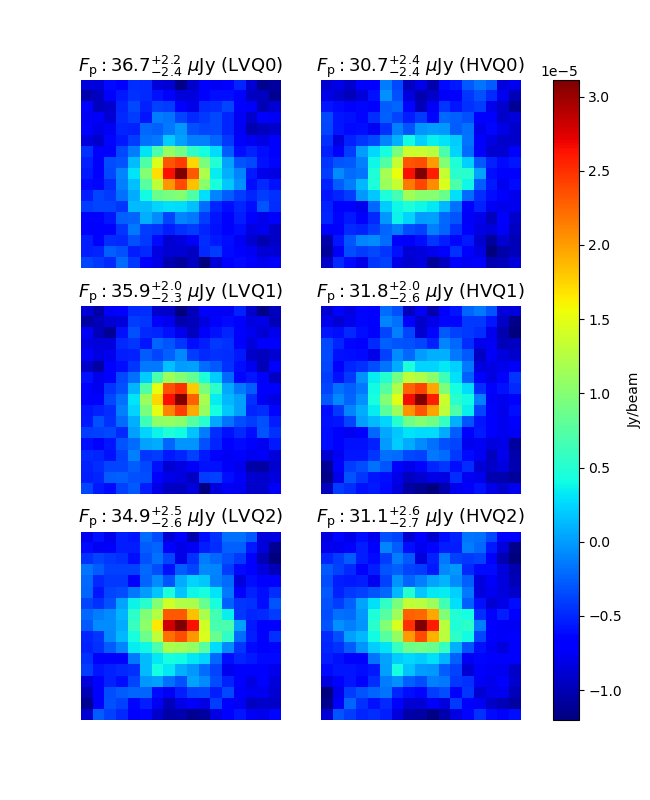}
	\includegraphics[width=3.4in]{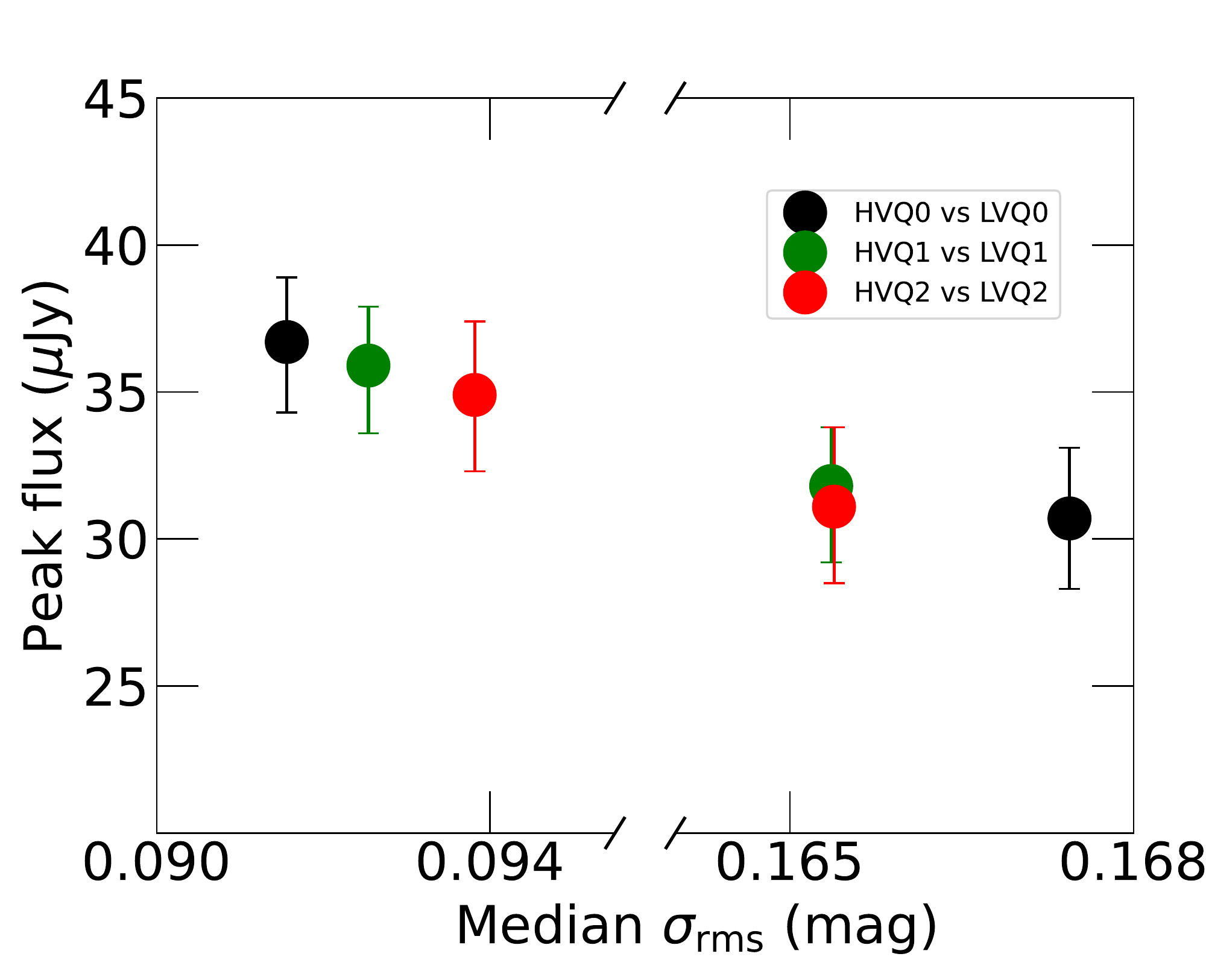}
	\caption{The stacked 1.4 GHz VLA images (upper, with a scale of $10\arcsec\ \times 10\arcsec$ and pixel size of 0.6\arcsec) of low- and high-variability quasars and derived median radio flux densities (lower), for quasars within the footprint of VLA-Stripe 82 which is considerably deeper than FIRST, but covering a smaller sky area. 
\label{radio_comparison1}}
\end{figure}

SDSS Stripe 82 is also covered by the 1.4 GHz VLA survey of the SDSS Southern Equatorial Stripe \citep[VLA-Stripe 82,][]{2011AJ....142....3H}, with an angular resolution of 1.8\arcsec\ and a rms sensitivity of $\sim$ 50 $\mu$Jy over 92 $\rm deg^{2}$. The VLA-Stripe 82 provides us considerably deeper radio images with better resolution, but for a smaller sample of quasars. 
Among the 3799 quasars within the footprint of VLA-Stripe 82, 280 ($\sim$7.4\%) have radio detections.
Cross-matching the 3799 quasars with \cite{2011ApJS..194...45S} to obtain $M_{\rm BH}$, $L_{\rm bol}$, $R_{\rm edd}$ and excluding sources with $<$ 20 epochs in SDSS $g$ band light curves, we obtain a sample of 3669 quasars. 
We repeat our sample splitting, and radio imaging stacking analyses using radio images from VLA-Stripe 82 for this sample. The results are presented in Fig. \ref{radio_comparison1}, with the derived median radio flux densities well consistent (within statistical errors) with those obtained using FIRST images (Fig. \ref{stacked_image} and \ref{bins5}). 


\begin{figure}
	\includegraphics[width=3.4in]{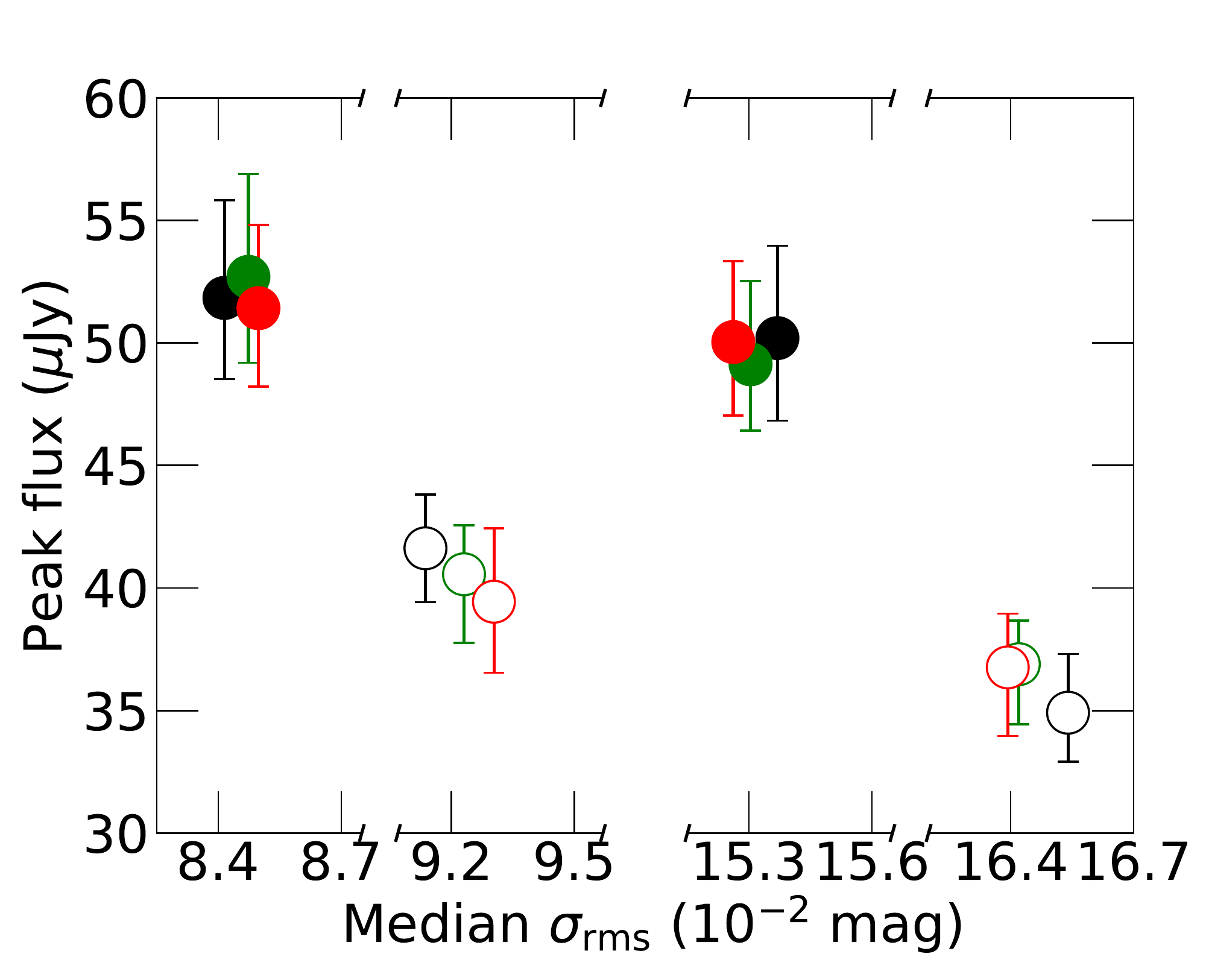}
	\caption{Similar to Fig. \ref{bins5} and \ref{radio_comparison1}, but for quasars which can be safely classified as radio-quiet based on radio fluxes or upper limits. Solid and open symbols (with colors same to Fig. \ref{bins5}) are derived using FIRST and VLA-Stripe82 images, respectively. 
	\label{Pure radio-quiet quasars}}
\end{figure}

In above analyses we have adopted the 45th percentile while stacking radio images to minimize the contamination by the $\sim$ 10\% radio-loud sources in the quasar sample. Alternatively and more conservatively, we could use only quasars which can be safely classified as radio-quiet based on available radio fluxes or upper limits, and use such sample for analyses. 
We adopt an upper limit of 1 mJy at 1.4 GHz for FIRST non-detections, and 0.3 mJy for VLA-Stripe 82 non-detections. 
We utilize the rest-frame 2500 \AA~ flux from \cite{2011ApJS..194...45S}, and assume a radio spectra index of $\alpha$ = 0.5 to estimate the radio-loudness $f_{\rm 5GHz}/f_{\rm 2500A}$ (with a threshold of $<$ 10 for radio-quiet sources).
We find 5476 out of 8819 quasars with FIRST coverage could be safely classified as radio-quiet, including 106 with FIRST detections. 
With deeper VLA-Stripe82 images, 3410 out 3669 quasars remain, including 71 radio detections.
Repeating our analyses (sample splitting and radio image stacking) using these samples, we present the simple median flux densities for low- and high-variability subsamples in Fig. \ref{Pure radio-quiet quasars}.
Clearly, we see similar trends that HVQs have tentatively weaker radio emission compared with LVQs. Not the absolute flux densities in Fig. \ref{Pure radio-quiet quasars} from the FIRST sample are considerably higher than those plotted in Fig. \ref{bins5} and \ref{radio_comparison1}. This is mainly because to ensure the radio quietness based on radio flux upper limit, the quasars we kept need to have higher luminosity and/or lower redshifts, thus higher average radio fluxes. For the VLA-Stripe82 sample, 93\% sources are kept as radio-quiet as VLA-Stripe82 images are sufficiently, thus the median fluxes plotted in Fig. \ref{Pure radio-quiet quasars} are somehow effectively 46.5th percentile, thus only slightly higher than 45th percentile. 

In summary, we find high-variability RQQs (with matched redshift, bolometric luminosity and SMBH mass) have tentatively weaker radio emission compared with low-variability ones.

\section{Discussion}
\subsection{The origin of radio emission in RQQs}
\cite{2018ApJ...868...58K} investigated the relation between UV/optical variation amplitude ($\sigma_{\rm rms}$) and X-ray loudness ($L_{\rm X}/L_{\rm bol}$) using a sample of X-ray detected quasars in Stripe 82X \citep{Ananna2017}. They found that quasars with more intense X-ray radiation (controlling the effects of other physical parameters, i.e., $z$, $M_{\rm BH}$, $L_{\rm bol}$ and $R_{\rm edd}$) are more variable in the UV/optical. It can be interpreted that the UV/optical variations and X-ray emission are both associated with magnetic turbulence, i.e., the stronger disk turbulence, the more energy can be dissipated into the corona which could be heated through magnetic reconnection. 
We divide the sample of \cite{2018ApJ...868...58K} into high- and low-variability quasars as we built HVQ1 and LVQ1 in this work. We find that while the HVQ1 sample of \cite{2018ApJ...868...58K} is more variable than LVQ1 by a factor of 1.80 (which is similar to the difference between our LVQ1/HVQ1 and LVQ2/HVQ2), their HVQ1 are X-ray brighter on average by a factor of 1.5 than LVQ1.

\begin{figure*}
	\includegraphics[width=3.4in]{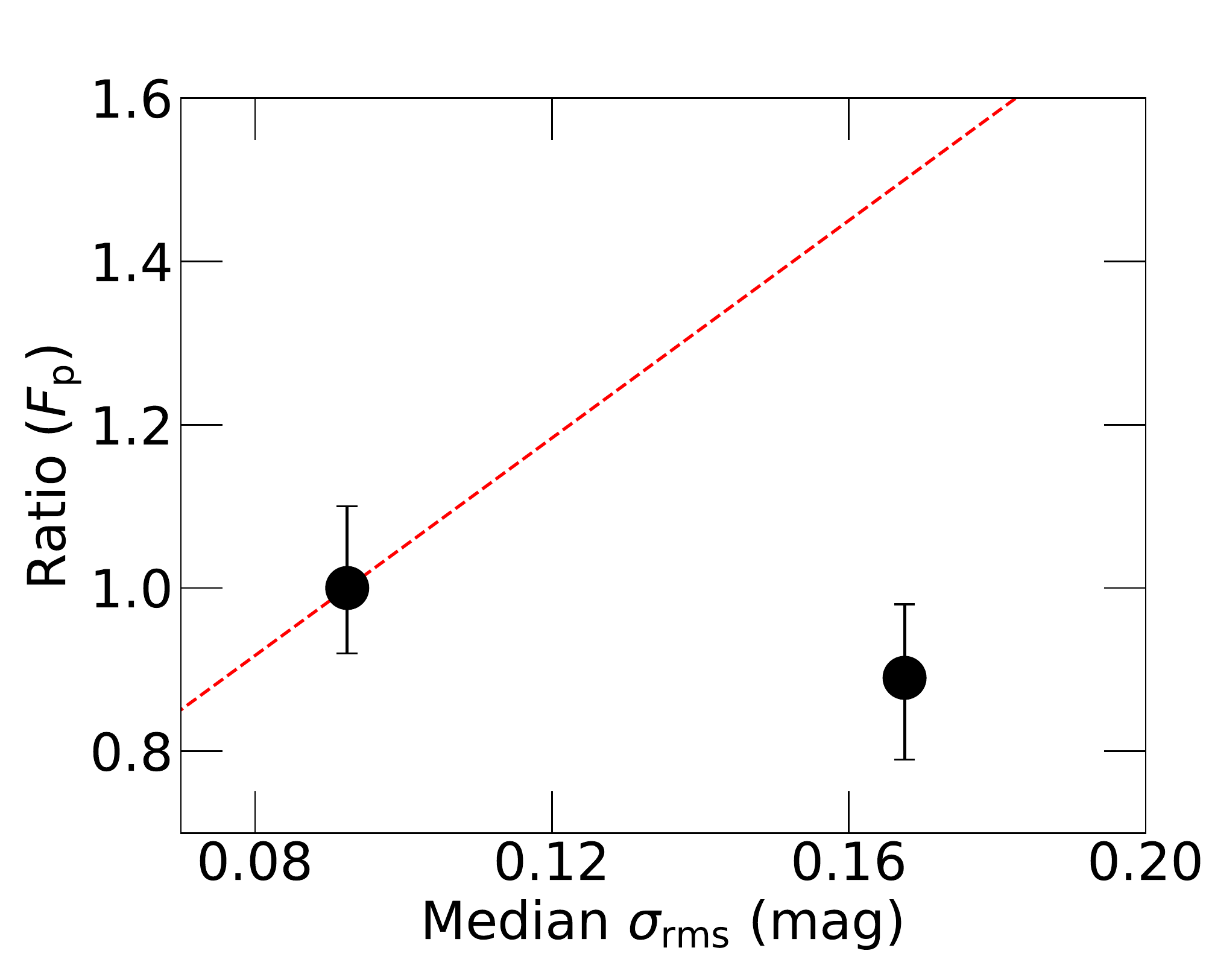}
	\includegraphics[width=3.4in]{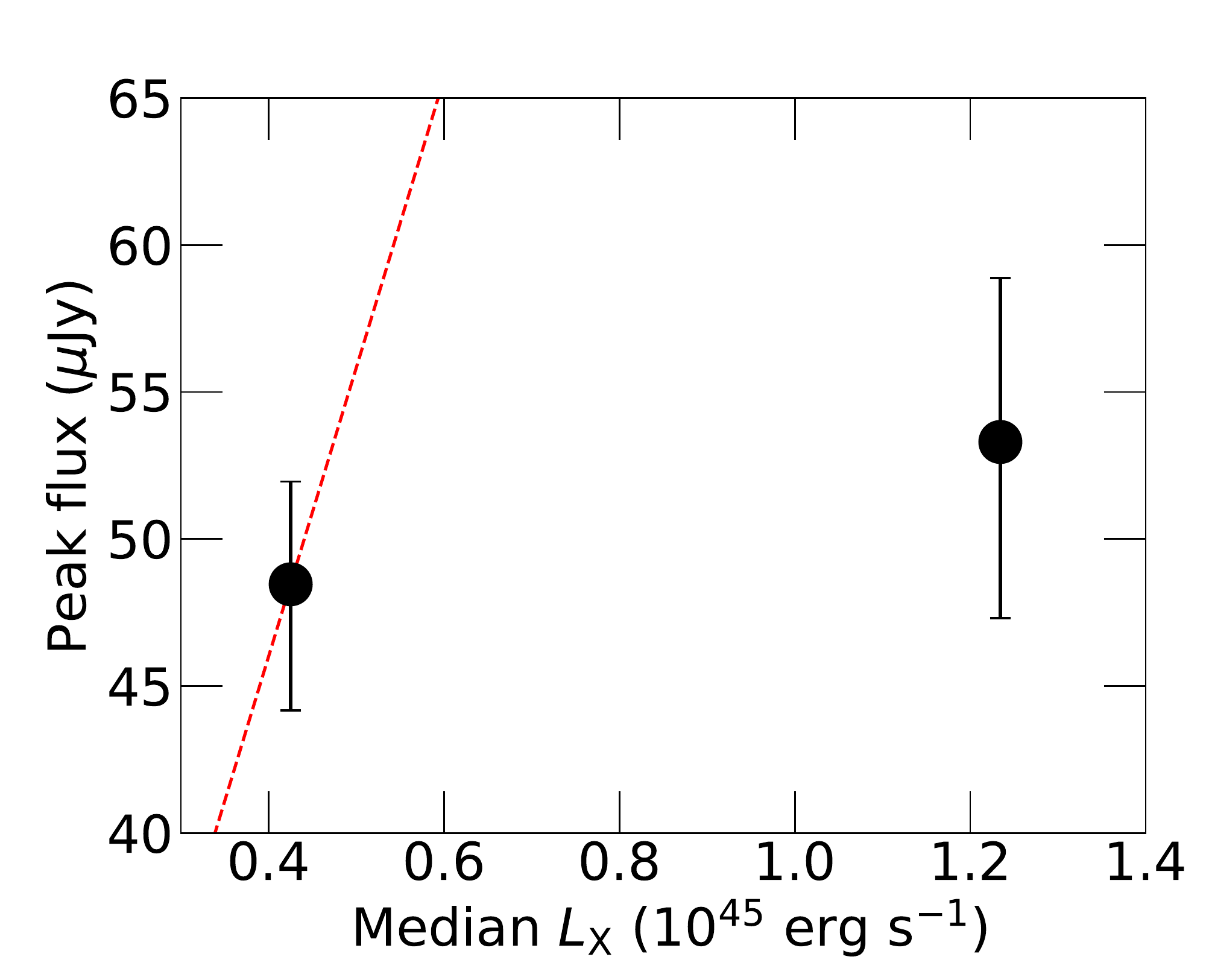}
	\caption{Left: comparison between our results (the median radio flux density ratios, i.e., scaled to LVQ2, versus $\sigma_{\rm rms}$ of our HVQ2 and LVQ2, black dots), and the prediction relation (red dashed line) based on the measured relation between X-ray loudness and $\sigma_{\rm rms}$ from Kang et al. (2018) and assuming a linear radio/X-ray relation. Right: the detected median radio flux densities versus median X-ray luminosity of two X-ray detected RQQ samples (black dots); the red dashed line mark the linear radio/X-ray relation. \label{radio_comparison}}
\end{figure*}

The radio emission is expected to linearly correlate with X-ray according to the relation between X-ray and radio emission with $L_{\rm R}$ $=$ $10^{-5}$ $L_{\rm X}$ in RQQs reported in \cite{2008MNRAS.390..847L}, probably driven by the Neupert effect (where $L_{\rm R}$ = d$L_{\rm X}/$d${t}$ as detected in stellar coronae, \citealt{Neupert1968,Guedel1993}). If so, we would expect our HVQ2 have radio emission $\sim$ 1.5 times stronger than LVQ2. 
However, our results 
show that the radio emission of HVQs are consistent with that of LVQs,
contrary to the prediction (see the left panel in Fig. \ref{radio_comparison}).

Our discovery thus indicates the 1.4 GHz FIRST radio emission of RQQs is unlikely driven by the central corona, or an extended region powered by the central corona. It would be intriguing to explore in the future the connection between UV/optical variability (thus X-ray loudness) and radio emission at higher frequencies, where the contribution of the corona to radio emission could be more significant.

On the other hand, if the radio emission in RQQs is dominated by weak jet, our discovery suggests that such jet should be driven by processes other than those which drive the the UV/optical variability and corona heating.
Note \cite{2019SCPMA..6269511C} suggested that
the inner accretion disk of RLQs fluctuates less compared with RQ ones, likely because the stronger magnetic field in RLQs could stabilize the inner disc,  i.e., implying a negative link between jet power and UV/optical variability.
In this case, the tentatively negative relation between UV/optical variability and radio emission in RQQs reported in this work may support the scenario that the radio emission in RQQs is dominated by weak jet.
Future studies of larger RQQ samples may reveal whether the tentatively negative relation is statistically robust, providing new clues to the radio emission origin.

The large-scale radio emission from star formation in the host galaxies may also fit our results (i.e., no statistically significant difference in radio emission between low- and high-variability quasars), as the accretion disk variations caused by random fluctuation may be no relevance with the host properties due to the origin of radio emission at spatial scales different by a factor of several orders. 
The scenario of star formation origin in RQQs is supported by recent low frequency radio observations. With Low Frequency Array \citep[LOFAR;][]{2013A&A...556A...2V} Two-metre Sky Survey \citep[LoTSS DR1;][]{2017A&A...598A.104S}, \cite{2019A&A...622A..11G}  found that the 144 MHz continuum emission of RQQs (quasars with 0.1<$L_{144\rm{MHz}}$/$L_{i\rm{band}}$<100) is consistent with being dominated by star formation based on the far-IR to radio correlation measured in local star-forming galaxies (but note the far-IR to radio correlation may evolves with redshift). 
However, the weak jet contribution to 144 MHz radio emission in RQQs might also be non-negligible. Through modeling the LoTSS radio luminosity distribution of quasars assuming both jet and star formation contribute to radio emission in every quasar, \cite{2021MNRAS.506.5888M} showed that the radio emission from jets is contributing down at least to the level comparable to the radio emission from star formation.

To summarize, we find more variable RQQs (at matched redshift, luminosity and SMBH mass) tend to have tentatively weaker radio emission, disfavoring the corona origin of radio radiation in RQQs. If the tentatively negative relation between UV/optical variability and radio emission in RQQs could be confirmed in the future, the weak jet origin would be preferred (or at least with significant contribution), as radio emission from star formation is expected to be irrelevant to the short term disc turbulence in quasars.

\subsection{Why no directly explore the connection between X-ray and radio emission in RQQs?}

We have shown that RQQs with stronger UV/optical variability, which are known to have relatively stronger X-ray emission, do not show stronger radio emission. This poses clear challenge to the corona origin of the radio emission in RQQs.
One might ask a simple question: why not directly explore the connection between X-ray and radio emission in RQQs?

First of all, the direct correlation between X-ray and radio power in RQQs may not yield useful constraints to the radio emission origin.
Numerous previous statistical studies have revealed correlations between radio and X-ray emission in RQQs and RQ AGN \citep[e.g.,][]{2000A&A...356..445B,2004ApJ...600L..31S,2007A&A...467..519P,2008MNRAS.390..847L,Panessa2015}, seeming to support that their radio emission associates with the central engine of black hole activity instead of the star formation of host galaxies. 
This however does not necessarily support the corona origin of radio emission in radio-quiet AGN, because if the radio emission comes from jet, it is also expected to correlate with the bolometric power of AGN. Furthermore,
studies also revealed correlations between the star formation rate and AGN power in radio-quiet AGN \citep[e.g.][]{Sargent2010, Thean2001,2021MNRAS.506.5888M}, 
thus the correlation between radio emission and AGN power in radio-quiet sources does not necessarily imply an accretion related origin of the radio emission either.

What if we explore the relation between radio and X-ray emission in RQQs after controlling the effect of bolometric luminosity?
Intuitively, this could be more straightforward than exploring the relation between UV/optical variability and radio emission.
However, as we will show below, this approach is biased thus disfavored compared with the one adopted in this work.

For illustration, we explore the correlation between X-ray and radio emission in RQQs (controlling the effect of bolometric luminosity) with an X-ray detected SDSS quasar sample (without requiring variability measurement). We cross-match the fourth XMM-Newton serendipitous source catalogue (4XMM-DR9, \citealt{2020A&A...641A.136W}) with high SDSS DR7 quasar catalog of \cite{2011ApJS..194...45S} within 5$''$, 
resulting 4252 quasars with X-ray detection(s) where we only retain the one detection with the highest values of maximum-likelihood for those sources that have multiple detections. 4153 out of them fall in the coverage of FIRST survey, and they are further cross-matched with FIRST catalog within 5$''$ to exclude potential radio-loud quasars. We find and exclude 485 quasars with FIRST detections, i.e., a fraction of $\sim$ 11$\%$. 
We perform multiple linear regression to derive the dependence of X-ray (0.2$-$12 keV) luminosity on $z$, $M_{\rm BH}$, and $R_{\rm edd}$ (similar to Equation \ref{Eq-regression}).
We then divide the X-ray detected RQQs sample equally into
X-ray brighter
and X-ray-fainter 
subsamples, based on their residual X-ray luminosity with respect to the expected values derived from the best-fit regression. The K-S test results demonstrate that these two subsamples already have statistically indistinguishable distributions of $z$, $M_{\rm BH}$, $L_{\rm bol}$ and $R_{\rm edd}$.
We find median peak flux densities 53.3$_{-6.0}^{+5.6}$ $\mu$Jy and 48.5$_{-4.3}^{+3.5}$ $\mu$Jy in X-ray brighter and X-ray-fainter subsamples\footnote{Here, the peak flux densities is directly from the median (50th percentile) of the quasars after excluding the $\sim 11\%$ radio detected ones.}, respectively, also contrary to a linear correlation between X-ray and radio emission (see the right panel of Fig. \ref{radio_comparison}).

But note that the X-ray brighter sample here yield slightly higher radio emission compared with the X-ray fainter sample, while our HVQ sample which is X-ray brighter than LVQ sample exhibits slightly lower radio emission than LVQ, although neither the differences is statistically significant. The subtle inconsistency between the two approaches could be due the uncertainties in bolometric luminosity or Eddington ratio. This is because the observed bolometric luminosity of an individual source could significantly deviate from the intrinsic value because of flux variability or inclination effect of the accretion disc emission, and since both radio and X-ray power correlates with the central engine, the uncertainties in bolometric luminosity could lead artificial partial correlations between radio and X-ray emission \citep[e.g.][]{2018ApJ...868...58K}. Though the results from two approaches are generally consistent, as the dependence of variability on bolometric luminosity is much weaker than that of X-ray or radio power \citep[e.g., see Fig. 2 in][]{2018ApJ...868...58K}, the primary approach presented in this work provides less biased results and thus is preferred.

\section*{Acknowledgements}
We thank the anonymous referee for constructive suggestions which is great helpful in improving the manuscript.
We thank Richard L. White, Xiaofeng Li, Mladen Novak, Zhengyi Cai, Lu Shen, Zesen Lin for useful discussions. This work was supported by the National Science Foundation of China (Grants No. 1890693, 12033006 $\&$ 12192221), the China Postdoctoral Science Foundation (Grant No. 2021M693089), and the CASFrontier Science Key Research Program (QYZDJ-SSW-SLH006). MHZ is supported by the Science and Technology Project funded by the Education Department of Jiangxi Province in China (Grant No. GJJ211733), and the Doctoral Scientific Research Foundation of Shangrao Normal University (Grant No. K6000449)

\section*{Data Availability}
This work is based on public images available from the FIRST and Stripe82 VLA archive.



\bibliographystyle{mnras}
\bibliography{ms} 

\begin{thebibliography}{}
\makeatletter
\relax
\def\mn@urlcharsother{\let\do\@makeother \do\$\do\&\do\#\do\^\do\_\do\%\do\~}
\def\mn@doi{\begingroup\mn@urlcharsother \@ifnextchar [ {\mn@doi@}
  {\mn@doi@[]}}
\def\mn@doi@[#1]#2{\def\@tempa{#1}\ifx\@tempa\@empty \href
  {http://dx.doi.org/#2} {doi:#2}\else \href {http://dx.doi.org/#2} {#1}\fi
  \endgroup}
\def\mn@eprint#1#2{\mn@eprint@#1:#2::\@nil}
\def\mn@eprint@arXiv#1{\href {http://arxiv.org/abs/#1} {{\tt arXiv:#1}}}
\def\mn@eprint@dblp#1{\href {http://dblp.uni-trier.de/rec/bibtex/#1.xml}
  {dblp:#1}}
\def\mn@eprint@#1:#2:#3:#4\@nil{\def\@tempa {#1}\def\@tempb {#2}\def\@tempc
  {#3}\ifx \@tempc \@empty \let \@tempc \@tempb \let \@tempb \@tempa \fi \ifx
  \@tempb \@empty \def\@tempb {arXiv}\fi \@ifundefined
  {mn@eprint@\@tempb}{\@tempb:\@tempc}{\expandafter \expandafter \csname
  mn@eprint@\@tempb\endcsname \expandafter{\@tempc}}}

\bibitem[\protect\citeauthoryear{{Ananna} et~al.,}{{Ananna}
  et~al.}{2017}]{Ananna2017}
{Ananna} T.~T.,  et~al., 2017, \apj, 850, 66

\bibitem[\protect\citeauthoryear{{Baldi}, {Laor}, {Behar}, {Horesh}, {Panessa},
  {McHardy}  \& {Kimball}}{{Baldi} et~al.}{2021}]{Baldi2021}
{Baldi} R.~D.,  {Laor} A.,  {Behar} E.,  {Horesh} A.,  {Panessa} F.,  {McHardy}
  I.,   {Kimball} A.,  2021, arXiv e-prints, p. arXiv:2107.14490

\bibitem[\protect\citeauthoryear{{Becker}, {White}  \& {Helfand}}{{Becker}
  et~al.}{1995}]{1995ApJ...450..559B}
{Becker} R.~H.,  {White} R.~L.,   {Helfand} D.~J.,  1995, \apj, 450, 559

\bibitem[\protect\citeauthoryear{{Behar}, {Vogel}, {Baldi}, {Smith}  \&
  {Mushotzky}}{{Behar} et~al.}{2018}]{2018MNRAS.478..399B}
{Behar} E.,  {Vogel} S.,  {Baldi} R.~D.,  {Smith} K.~L.,   {Mushotzky} R.~F.,
  2018, \mnras, 478, 399

\bibitem[\protect\citeauthoryear{{Blandford} \& {K{\"o}nigl}}{{Blandford} \&
  {K{\"o}nigl}}{1979}]{1979ApJ...232...34B}
{Blandford} R.~D.,  {K{\"o}nigl} A.,  1979, \apj, 232, 34

\bibitem[\protect\citeauthoryear{{Blandford}, {Meier}  \&
  {Readhead}}{{Blandford} et~al.}{2019}]{2019ARA&A..57..467B}
{Blandford} R.,  {Meier} D.,   {Readhead} A.,  2019, \araa, 57, 467

\bibitem[\protect\citeauthoryear{{Blundell} \& {Beasley}}{{Blundell} \&
  {Beasley}}{1998}]{1998MNRAS.299..165B}
{Blundell} K.~M.,  {Beasley} A.~J.,  1998, \mnras, 299, 165

\bibitem[\protect\citeauthoryear{{Blundell} \& {Rawlings}}{{Blundell} \&
  {Rawlings}}{2001}]{2001ApJ...562L...5B}
{Blundell} K.~M.,  {Rawlings} S.,  2001, \apjl, 562, L5

\bibitem[\protect\citeauthoryear{{Blundell}, {Beasley}  \&
  {Bicknell}}{{Blundell} et~al.}{2003}]{2003ApJ...591L.103B}
{Blundell} K.~M.,  {Beasley} A.~J.,   {Bicknell} G.~V.,  2003, \apjl, 591, L103

\bibitem[\protect\citeauthoryear{{Bonzini} et~al.,}{{Bonzini}
  et~al.}{2015}]{2015MNRAS.453.1079B}
{Bonzini} M.,  et~al., 2015, \mnras, 453, 1079

\bibitem[\protect\citeauthoryear{{Brinkmann}, {Laurent-Muehleisen}, {Voges},
  {Siebert}, {Becker}, {Brotherton}, {White}  \& {Gregg}}{{Brinkmann}
  et~al.}{2000}]{2000A&A...356..445B}
{Brinkmann} W.,  {Laurent-Muehleisen} S.~A.,  {Voges} W.,  {Siebert} J.,
  {Becker} R.~H.,  {Brotherton} M.~S.,  {White} R.~L.,   {Gregg} M.~D.,  2000,
  \aap, \href {https://ui.adsabs.harvard.edu/abs/2000A&A...356..445B} {356,
  445}

\bibitem[\protect\citeauthoryear{{Cai}, {Wang}, {Gu}, {Sun}, {Wu}, {Huang}  \&
  {Chen}}{{Cai} et~al.}{2016}]{2016ApJ...826....7C}
{Cai} Z.-Y.,  {Wang} J.-X.,  {Gu} W.-M.,  {Sun} Y.-H.,  {Wu} M.-C.,  {Huang}
  X.-X.,   {Chen} X.-Y.,  2016, \apj, 826, 7

\bibitem[\protect\citeauthoryear{{Cai}, {Wang}, {Zhu}, {Sun}, {Gu}, {Cao}  \&
  {Yuan}}{{Cai} et~al.}{2018}]{2018ApJ...855..117C}
{Cai} Z.-Y.,  {Wang} J.-X.,  {Zhu} F.-F.,  {Sun} M.-Y.,  {Gu} W.-M.,  {Cao}
  X.-W.,   {Yuan} F.,  2018, \apj, 855, 117

\bibitem[\protect\citeauthoryear{{Cai}, {Sun}, {Wang}, {Zhu}, {Gu}  \&
  {Yuan}}{{Cai} et~al.}{2019}]{2019SCPMA..6269511C}
{Cai} Z.,  {Sun} Y.,  {Wang} J.,  {Zhu} F.,  {Gu} W.,   {Yuan} F.,  2019,
  Science China Physics, Mechanics, and Astronomy, 62, 69511

\bibitem[\protect\citeauthoryear{{Falcke} \& {Biermann}}{{Falcke} \&
  {Biermann}}{1995}]{1995A&A...293..665F}
{Falcke} H.,  {Biermann} P.~L.,  1995, \aap, 293, 665

\bibitem[\protect\citeauthoryear{{Garn} \& {Alexander}}{{Garn} \&
  {Alexander}}{2009}]{2009MNRAS.394..105G}
{Garn} T.,  {Alexander} P.,  2009, \mnras, 394, 105

\bibitem[\protect\citeauthoryear{{Guedel} \& {Benz}}{{Guedel} \&
  {Benz}}{1993}]{Guedel1993}
{Guedel} M.,  {Benz} A.~O.,  1993, \apjl, 405, L63

\bibitem[\protect\citeauthoryear{{G{\"u}rkan} et~al.,}{{G{\"u}rkan}
  et~al.}{2019}]{2019A&A...622A..11G}
{G{\"u}rkan} G.,  et~al., 2019, \aap, 622, A11

\bibitem[\protect\citeauthoryear{{Hardcastle} \& {Croston}}{{Hardcastle} \&
  {Croston}}{2020}]{2020NewAR..8801539H}
{Hardcastle} M.~J.,  {Croston} J.~H.,  2020, \nar, 88, 101539

\bibitem[\protect\citeauthoryear{{Helfand}, {White}  \& {Becker}}{{Helfand}
  et~al.}{2015}]{2015ApJ...801...26H}
{Helfand} D.~J.,  {White} R.~L.,   {Becker} R.~H.,  2015, \apj, 801, 26

\bibitem[\protect\citeauthoryear{{Hodge}, {Becker}, {White}, {Richards}  \&
  {Zeimann}}{{Hodge} et~al.}{2011}]{2011AJ....142....3H}
{Hodge} J.~A.,  {Becker} R.~H.,  {White} R.~L.,  {Richards} G.~T.,   {Zeimann}
  G.~R.,  2011, \aj, 142, 3

\bibitem[\protect\citeauthoryear{{Hwang}, {Zakamska}, {Alexandroff}, {Hamann},
  {Greene}, {Perrotta}  \& {Richards}}{{Hwang}
  et~al.}{2018}]{2018MNRAS.477..830H}
{Hwang} H.-C.,  {Zakamska} N.~L.,  {Alexandroff} R.~M.,  {Hamann} F.,  {Greene}
  J.~E.,  {Perrotta} S.,   {Richards} G.~T.,  2018, \mnras, 477, 830

\bibitem[\protect\citeauthoryear{{Inoue} \& {Doi}}{{Inoue} \&
  {Doi}}{2018}]{Inoue2018}
{Inoue} Y.,  {Doi} A.,  2018, \apj, 869, 114

\bibitem[\protect\citeauthoryear{{Ivezi{\'c}} et~al.,}{{Ivezi{\'c}}
  et~al.}{2002}]{2002AJ....124.2364I}
{Ivezi{\'c}} {\v{Z}}.,  et~al., 2002, \aj, 124, 2364

\bibitem[\protect\citeauthoryear{{Kang}, {Wang}, {Cai}, {Guo}, {Zhu}, {Cao},
  {Gu}  \& {Yuan}}{{Kang} et~al.}{2018}]{2018ApJ...868...58K}
{Kang} W.-y.,  {Wang} J.-X.,  {Cai} Z.-Y.,  {Guo} H.-X.,  {Zhu} F.-F.,  {Cao}
  X.-W.,  {Gu} W.-M.,   {Yuan} F.,  2018, \apj, 868, 58

\bibitem[\protect\citeauthoryear{{Kang}, {Wang}, {Cai}  \& {Ren}}{{Kang}
  et~al.}{2021}]{2021ApJ...911..148K}
{Kang} W.-Y.,  {Wang} J.-X.,  {Cai} Z.-Y.,   {Ren} W.-K.,  2021, \apj, 911, 148

\bibitem[\protect\citeauthoryear{{Karim} et~al.,}{{Karim}
  et~al.}{2011}]{2011ApJ...730...61K}
{Karim} A.,  et~al., 2011, \apj, 730, 61

\bibitem[\protect\citeauthoryear{{Kellermann}, {Sramek}, {Schmidt}, {Shaffer}
  \& {Green}}{{Kellermann} et~al.}{1989}]{1989AJ.....98.1195K}
{Kellermann} K.~I.,  {Sramek} R.,  {Schmidt} M.,  {Shaffer} D.~B.,   {Green}
  R.,  1989, \aj, 98, 1195

\bibitem[\protect\citeauthoryear{{Kellermann}, {Sramek}, {Schmidt}, {Green}  \&
  {Shaffer}}{{Kellermann} et~al.}{1994}]{1994AJ....108.1163K}
{Kellermann} K.~I.,  {Sramek} R.~A.,  {Schmidt} M.,  {Green} R.~F.,   {Shaffer}
  D.~B.,  1994, \aj, 108, 1163

\bibitem[\protect\citeauthoryear{{Kellermann}, {Condon}, {Kimball}, {Perley}
  \& {Ivezi{\'c}}}{{Kellermann} et~al.}{2016}]{2016ApJ...831..168K}
{Kellermann} K.~I.,  {Condon} J.~J.,  {Kimball} A.~E.,  {Perley} R.~A.,
  {Ivezi{\'c}} {\v{Z}}.,  2016, \apj, 831, 168

\bibitem[\protect\citeauthoryear{{Kelly}, {Bechtold}  \&
  {Siemiginowska}}{{Kelly} et~al.}{2009}]{2009ApJ...698..895K}
{Kelly} B.~C.,  {Bechtold} J.,   {Siemiginowska} A.,  2009, \apj, 698, 895

\bibitem[\protect\citeauthoryear{{Kukula}, {Dunlop}, {Hughes}  \&
  {Rawlings}}{{Kukula} et~al.}{1998}]{1998MNRAS.297..366K}
{Kukula} M.~J.,  {Dunlop} J.~S.,  {Hughes} D.~H.,   {Rawlings} S.,  1998,
  \mnras, 297, 366

\bibitem[\protect\citeauthoryear{{Laor} \& {Behar}}{{Laor} \&
  {Behar}}{2008}]{2008MNRAS.390..847L}
{Laor} A.,  {Behar} E.,  2008, \mnras, 390, 847

\bibitem[\protect\citeauthoryear{{Leipski}, {Falcke}, {Bennert}  \&
  {H{\"u}ttemeister}}{{Leipski} et~al.}{2006}]{2006A&A...455..161L}
{Leipski} C.,  {Falcke} H.,  {Bennert} N.,   {H{\"u}ttemeister} S.,  2006,
  \aap, 455, 161

\bibitem[\protect\citeauthoryear{{MacLeod} et~al.,}{{MacLeod}
  et~al.}{2012}]{2012ApJ...753..106M}
{MacLeod} C.~L.,  et~al., 2012, \apj, 753, 106

\bibitem[\protect\citeauthoryear{{Macfarlane} et~al.,}{{Macfarlane}
  et~al.}{2021}]{2021MNRAS.506.5888M}
{Macfarlane} C.,  et~al., 2021, \mnras, 506, 5888

\bibitem[\protect\citeauthoryear{{Miller}, {Rawlings}  \& {Saunders}}{{Miller}
  et~al.}{1993}]{1993MNRAS.263..425M}
{Miller} P.,  {Rawlings} S.,   {Saunders} R.,  1993, \mnras, 263, 425

\bibitem[\protect\citeauthoryear{{Mullaney}, {Alexander}, {Fine}, {Goulding},
  {Harrison}  \& {Hickox}}{{Mullaney} et~al.}{2013}]{2013MNRAS.433..622M}
{Mullaney} J.~R.,  {Alexander} D.~M.,  {Fine} S.,  {Goulding} A.~D.,
  {Harrison} C.~M.,   {Hickox} R.~C.,  2013, \mnras, 433, 622

\bibitem[\protect\citeauthoryear{{Neupert}}{{Neupert}}{1968}]{Neupert1968}
{Neupert} W.~M.,  1968, \apjl, 153, L59

\bibitem[\protect\citeauthoryear{{Olsson}, {Aalto}, {Thomasson}  \&
  {Beswick}}{{Olsson} et~al.}{2010}]{2010A&A...513A..11O}
{Olsson} E.,  {Aalto} S.,  {Thomasson} M.,   {Beswick} R.,  2010, \aap, 513,
  A11

\bibitem[\protect\citeauthoryear{{Padovani}}{{Padovani}}{2016}]{Padovani2016}
{Padovani} P.,  2016, \aapr, 24, 13

\bibitem[\protect\citeauthoryear{{Padovani}}{{Padovani}}{2017}]{Padovani2017NA}
{Padovani} P.,  2017, Nature Astronomy, 1, 0194

\bibitem[\protect\citeauthoryear{{Padovani} et~al.,}{{Padovani}
  et~al.}{2017}]{2017A&ARv..25....2P}
{Padovani} P.,  et~al., 2017, \aapr, 25, 2

\bibitem[\protect\citeauthoryear{{Panessa} \& {Giroletti}}{{Panessa} \&
  {Giroletti}}{2013}]{2013MNRAS.432.1138P}
{Panessa} F.,  {Giroletti} M.,  2013, \mnras, 432, 1138

\bibitem[\protect\citeauthoryear{{Panessa}, {Barcons}, {Bassani}, {Cappi},
  {Carrera}, {Ho}  \& {Pellegrini}}{{Panessa}
  et~al.}{2007}]{2007A&A...467..519P}
{Panessa} F.,  {Barcons} X.,  {Bassani} L.,  {Cappi} M.,  {Carrera} F.~J.,
  {Ho} L.~C.,   {Pellegrini} S.,  2007, \aap, 467, 519

\bibitem[\protect\citeauthoryear{{Panessa} et~al.,}{{Panessa}
  et~al.}{2015}]{Panessa2015}
{Panessa} F.,  et~al., 2015, \mnras, 447, 1289

\bibitem[\protect\citeauthoryear{{Panessa}, {Baldi}, {Laor}, {Padovani},
  {Behar}  \& {McHardy}}{{Panessa} et~al.}{2019}]{2019NatAs...3..387P}
{Panessa} F.,  {Baldi} R.~D.,  {Laor} A.,  {Padovani} P.,  {Behar} E.,
  {McHardy} I.,  2019, Nature Astronomy, 3, 387

\bibitem[\protect\citeauthoryear{{Radcliffe}, {Barthel}, {Garrett}, {Beswick},
  {Thomson}  \& {Muxlow}}{{Radcliffe} et~al.}{2021}]{2021A&A...649L...9R}
{Radcliffe} J.~F.,  {Barthel} P.~D.,  {Garrett} M.~A.,  {Beswick} R.~J.,
  {Thomson} A.~P.,   {Muxlow} T.~W.~B.,  2021, \aap, 649, L9

\bibitem[\protect\citeauthoryear{{Raginski} \& {Laor}}{{Raginski} \&
  {Laor}}{2016}]{2016MNRAS.459.2082R}
{Raginski} I.,  {Laor} A.,  2016, \mnras, 459, 2082

\bibitem[\protect\citeauthoryear{{Salvato}, {Greiner}  \&
  {Kuhlbrodt}}{{Salvato} et~al.}{2004}]{2004ApJ...600L..31S}
{Salvato} M.,  {Greiner} J.,   {Kuhlbrodt} B.,  2004, \apjl, 600, L31

\bibitem[\protect\citeauthoryear{{Sargent} et~al.,}{{Sargent}
  et~al.}{2010}]{Sargent2010}
{Sargent} M.~T.,  et~al., 2010, \apjs, 186, 341

\bibitem[\protect\citeauthoryear{{Sesar} et~al.,}{{Sesar}
  et~al.}{2007}]{2007AJ....134.2236S}
{Sesar} B.,  et~al., 2007, \aj, 134, 2236

\bibitem[\protect\citeauthoryear{{Shen} et~al.,}{{Shen}
  et~al.}{2011}]{2011ApJS..194...45S}
{Shen} Y.,  et~al., 2011, \apjs, 194, 45

\bibitem[\protect\citeauthoryear{{Shimwell} et~al.,}{{Shimwell}
  et~al.}{2017}]{2017A&A...598A.104S}
{Shimwell} T.~W.,  et~al., 2017, \aap, 598, A104

\bibitem[\protect\citeauthoryear{{Smith} et~al.,}{{Smith}
  et~al.}{2020}]{2020MNRAS.492.4216S}
{Smith} K.~L.,  et~al., 2020, \mnras, 492, 4216

\bibitem[\protect\citeauthoryear{{Thean}, {Pedlar}, {Kukula}, {Baum}  \&
  {O'Dea}}{{Thean} et~al.}{2001}]{Thean2001}
{Thean} A.,  {Pedlar} A.,  {Kukula} M.~J.,  {Baum} S.~A.,   {O'Dea} C.~P.,
  2001, \mnras, 325, 737

\bibitem[\protect\citeauthoryear{{Thompson}, {Moran}  \& {Swenson}}{{Thompson}
  et~al.}{2017}]{2017isra.book.....T}
{Thompson} A.~R.,  {Moran} J.~M.,   {Swenson} George~W. J.,  2017,
  {Interferometry and Synthesis in Radio Astronomy, 3rd Edition}

\bibitem[\protect\citeauthoryear{{Ulvestad}, {Antonucci}  \&
  {Barvainis}}{{Ulvestad} et~al.}{2005}]{2005ApJ...621..123U}
{Ulvestad} J.~S.,  {Antonucci} R. R.~J.,   {Barvainis} R.,  2005, \apj, 621,
  123

\bibitem[\protect\citeauthoryear{{Wang}, {An}, {Jaiswal}, {Mohan}, {Wang},
  {Baan}, {Zhang}  \& {Yang}}{{Wang} et~al.}{2021}]{2021MNRAS.504.3823W}
{Wang} A.,  {An} T.,  {Jaiswal} S.,  {Mohan} P.,  {Wang} Y.,  {Baan} W.~A.,
  {Zhang} Y.,   {Yang} X.,  2021, \mnras, 504, 3823

\bibitem[\protect\citeauthoryear{{Webb} et~al.,}{{Webb}
  et~al.}{2020}]{2020A&A...641A.136W}
{Webb} N.~A.,  et~al., 2020, \aap, 641, A136

\bibitem[\protect\citeauthoryear{{White}, {Helfand}, {Becker}, {Glikman}  \&
  {de Vries}}{{White} et~al.}{2007}]{2007ApJ...654...99W}
{White} R.~L.,  {Helfand} D.~J.,  {Becker} R.~H.,  {Glikman} E.,   {de Vries}
  W.,  2007, \apj, 654, 99

\bibitem[\protect\citeauthoryear{{White}, {Jarvis}, {Kalfountzou},
  {Hardcastle}, {Verma}, {Cao Orjales}  \& {Stevens}}{{White}
  et~al.}{2017}]{White2017}
{White} S.~V.,  {Jarvis} M.~J.,  {Kalfountzou} E.,  {Hardcastle} M.~J.,
  {Verma} A.,  {Cao Orjales} J.~M.,   {Stevens} J.,  2017, \mnras, 468, 217

\bibitem[\protect\citeauthoryear{{Zakamska} \& {Greene}}{{Zakamska} \&
  {Greene}}{2014}]{2014MNRAS.442..784Z}
{Zakamska} N.~L.,  {Greene} J.~E.,  2014, \mnras, 442, 784

\bibitem[\protect\citeauthoryear{{Zakamska} et~al.,}{{Zakamska}
  et~al.}{2016}]{2016MNRAS.455.4191Z}
{Zakamska} N.~L.,  et~al., 2016, \mnras, 455, 4191

\bibitem[\protect\citeauthoryear{{Zhou} \& {Gu}}{{Zhou} \&
  {Gu}}{2021}]{2021RAA....21....4Z}
{Zhou} M.-H.,  {Gu} M.-F.,  2021, Research in Astronomy and Astrophysics, 21,
  004

\bibitem[\protect\citeauthoryear{{Zuo}, {Wu}, {Liu}  \& {Jiao}}{{Zuo}
  et~al.}{2012}]{2012ApJ...758..104Z}
{Zuo} W.,  {Wu} X.-B.,  {Liu} Y.-Q.,   {Jiao} C.-L.,  2012, \apj, 758, 104

\bibitem[\protect\citeauthoryear{{van Haarlem} et~al.,}{{van Haarlem}
  et~al.}{2013}]{2013A&A...556A...2V}
{van Haarlem} M.~P.,  et~al., 2013, \aap, 556, A2

\bibitem[\protect\citeauthoryear{{van der Walt}, {Colbert}  \&
  {Varoquaux}}{{van der Walt} et~al.}{2011}]{2011CSE....13b..22V}
{van der Walt} S.,  {Colbert} S.~C.,   {Varoquaux} G.,  2011, Computing in
  Science and Engineering, 13, 22

\makeatother
\end{thebibliography}




\bsp	
\label{lastpage}
\end{document}